\documentclass[12pt]{article}
\usepackage[left=2cm, right=2cm,top=2cm, bottom=2cm]{geometry}
\usepackage{graphicx}
\usepackage{float}
\usepackage{color}
\usepackage{authblk}
\usepackage{amssymb,amsfonts,amsmath}

\usepackage{graphicx}
\usepackage{epstopdf}
\usepackage[extension=xxx]{hyperref}
\newcommand{\unit}[1]{\ensuremath{\, \mathrm{#1}}}

\usepackage{color}

\def\be{\begin{equation}}   \def\ee{\end{equation}}
\def\eq#1{{Eq~(\ref{#1})}}    \def\fig#1{{Fig.\ref{#1}}}

\begin{document}
\title{Stochastic Model of Maturation and Vesicular Exchange in Cellular Organelles\\ \vspace{0.5cm} {\large Running title: Maturation and Exchange in Organelles}}
\date{\today}
\author{Quentin Vagne and Pierre Sens}
\affil[1]{Institut Curie, PSL Research University, CNRS, UMR 168, 23 rue d'Ulm, F-75005, Paris, France.}
\maketitle

\section*{Abstract}
The dynamical organisation of membrane-bound organelles along intracellular transport pathways relies on vesicular exchange between organelles and on the maturation of the organelle's composition by enzymatic reactions or exchange with the cytoplasm. The relative importance of each mechanism in controlling organelle dynamics remains controversial, in particular for transport through the Golgi apparatus. Using a stochastic model, we identify two classes of dynamical behaviour that can lead to full maturation of membrane-bound compartments. In the first class, maturation corresponds to the stochastic escape from a steady-state in which export is dominated by vesicular exchange, and is very unlikely for large compartments. In the second class, it occurs in a quasi-deterministic fashion and is  almost size independent. Whether a system belongs to the first or second class is largely controlled by homotypic fusion.

\newpage

\section*{Introduction}

The hallmark of eukaryotic cells is their compartmentalization into specialized  organelles defining different biochemical environments within the cell. These compartments are bounded by a fluid lipid membrane and are highly dynamical, constantly exchanging components with one another through the budding and fusion of small transport vesicles \cite{bonifacino:2004}.  
The presence of different combinations of lipids and proteins in the membrane of different organelles defines different membrane ``identities''  and direct vesicular exchange by controlling the activity of membrane-associated proteins, such as coat proteins that drive vesicle budding, and  tethers and SNAREs that control vesicle fusion \cite{munro:2004,munro:2005}. 
This is required for the existence of well-defined intracellular transport pathways, such as the endocytic pathways, from the cell plasma membrane to early endosomes and late endosomes \cite{vandergoot:2006}, and the secretory pathway, from the endoplasmic reticulum (ER) to the Golgi apparatus and the trans-Golgi network  \cite{kelly1985pathways}.

Members of the Rab GTPase family play an important role in defining the membrane identity and regulate all steps of membrane traffic \cite{schimmoller:1998,zerial:2001,stenmark:2001,stenmark:2009}. In particular, Rabs are involved in homotypic fusion,  the propensity of a vesicle to fuse with a compartment of similar identity, a process relevant for the spatio-temporal organisation of both the endosomal network \cite{rink:2005} and the Golgi apparatus \cite{pfeffer:2010b}. Remarkably, the membrane identity of an organelle can change over time in a process called maturation. Maturation has been observed in the endosomal network, where Rab5 positive early endosomes mature into Rab7 positive late endosomes over a time scale of order $10 \unit{min}$ \cite{rink:2005}. It has also been observed in the Golgi apparatus of \textit{Saccharomyces cerevisiae}. In this organism,  the Golgi sub-compartments, called cisternae,  are dispersed throughout the cytoplasm and can be seen to mature from a {\em cis} ({\em i.e.} early) to a {\em trans}  ({\em i.e.} late) identity in about $1-2\unit{min}$ \cite{Matsuura:2006,losev:2006}. 

\begin{figure}[t]
\centering{\includegraphics[width=3.25in]{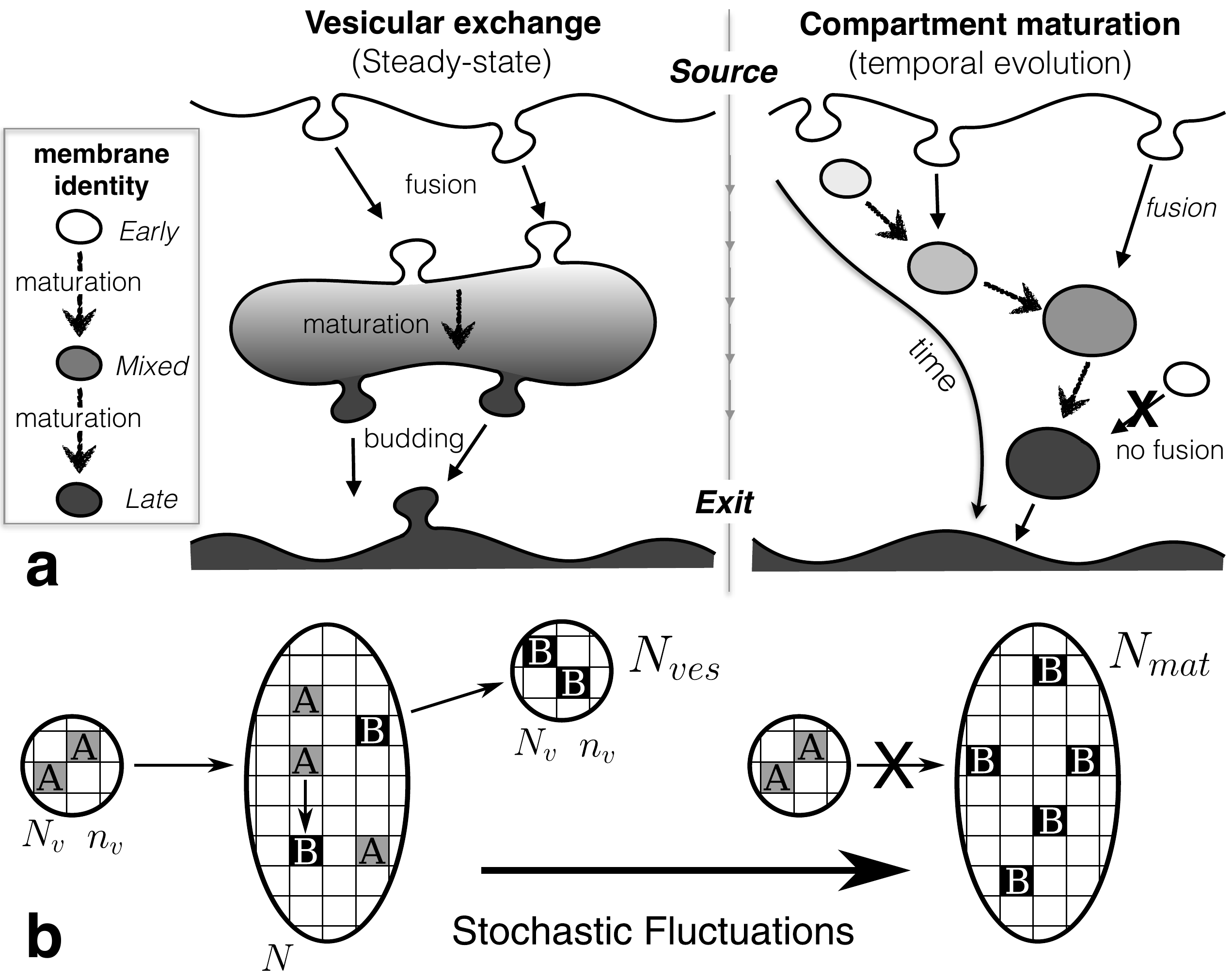}}
\caption{\label{fig:sketch} {\bf a}. Sketch of the possible dynamics of an organelle receiving a vesicular influx of early (white) membrane identity undergoing maturation into late (dark) identity inside the organelle and being exported by vesicular budding. The organelle could represent an early endosome, for which the source is the plasma membrane and the exit the pool of late endosomes, or the Golgi apparatus, for which the source is the Endoplasmic Reticulum and the exit the Trans Golgi Network. The organelle can be at steady-state (``vesicular exchange'' - left) where the influx of immature vesicles is balanced by an outflux of mature vesicles, or show a progressive evolution from an early to a late identity (``compartment maturation'' - right), where the influx is balanced by an outflux of mature compartments. {\bf b}. Sketches of the theoretical model: incoming and outgoing vesicles contain $N_{v}$ sites including $n_{v}$ immature ($A$ - grey) or mature ($B$ - black) components, respectively. The fusion of incoming vesicles requires the presence of $A$ sites in the compartment. Stochastic fluctuations lead to  full   maturation and the  isolation of the compartment from the influx, after what a new compartment is created {\em de novo}. 
}
\end{figure}

Vesicular export from early to late compartments and the biochemical maturation of early compartments into late compartments constitute  two distinct mechanisms allowing progression along secretory and endocytic pathways, raising  a fundamental question as to their relative importance for intracellular trafficking. In other words, \textit{are organelles  steady-state structures receiving, processing, and exporting transiting cargoes, or are they transient structures that are nucleated by an incoming flux and undergo full maturation} (see \fig{fig:sketch} for a sketch).
This question is particularly debated for the Golgi apparatus. In most animal and plant cells, it is made of individual compartment (cisternae), stacked together in a polarized way with an entry ({\em cis}) face and an exit ({\em trans}) face. Whether Golgi transport occurs by inter-cisternal {\em vesicular exchange} or by full {\em cisternal maturation} is still highly controversial \cite{emr:2009}. This question is of high physiological relevance considering the involvement of Golgi dysfunction in many pathologies, including Alzheimer and cancers \cite{wu:2002,kellokumpu:2002,gonatas2006fragmentation,hu:2007b,bexiga:2013}.  

Beyond the case of the Golgi apparatus, the interplay between biochemical maturation and vesicular exchange in cellular transport pathways is an  issue relevant for many aspects of intracellular organisation, and we currently lack a quantitative framework to address it.
Several physical models of intracellular transport have been developed in recent years \cite{Heinrich:2005,binder:2009,dmitrieff:2011,foret:2012}. These studies generally focus on steady-state properties, and the inherently stochastic nature of intracellular transport has been much less explored \cite{gong:2010,bressloff:2013}. Stochasticity should however play an important role, as the fusion/budding of a few tens of  vesicles (of diameter $\sim50-100\unit{nm}$), is enough to completely renew  the membrane composition of an endosome or a Golgi cisterna (of area $\sim0.2-0.5 \unit{\mu m^2}$). This explains the strong fluctuation of the size and composition of early endosomes \cite{rink:2005}. 

 We investigate theoretically a stochastic, one-compartment model (sketched in \fig{fig:sketch}), that includes both aspect of organelle dynamics. Immature membrane components are injected by the fusion of incoming vesicles, undergo biochemical maturation, and are exported by vesicular budding. We  precisely quantify whether the outflux of mature components  predominantly occurs by vesicular export from a steady-state compartment of fixed biochemical identities or by the full maturation of the entire compartment. 
Our model, which is investigated both analytically and with numerical simulations, is an application of the theory of birth-and-death stochastic processes, which are used to great extent in many areas of biology \cite{novozhilov:2006} and population dynamics \cite{assaf:2010}. It
establishes the importance of stochasticity in controlling the balance between  vesicular exchange and compartment maturation and identifies the key control parameters as being the ratio of vesicle injection to budding rate, and the ratio of biochemical maturation to budding rates.

\section*{Method}

We consider a membrane-bound compartment receiving a vesicular influx of components of a given (early) identity called $A$ which, after being converted  into a late identity $B$ by a maturation process, exit the compartment by selective vesicle budding (\fig{fig:sketch} b). 
At steady-state, the vesicular influx of immature components is entirely converted into vesicular out-flux of mature components, which corresponds to the  {\em vesicular exchange} mechanism. In practice however, one expects that stochastic fluctuations around the steady-state, which will be comparatively more important for small compartments,  eventually lead to the full maturation of all $A$ sites into the $B$ identity. Homotypic fusion makes it unlikely that a vesicle of immature identity $A$ will fuse with a fully mature compartment of $B$ components. To account for this, the vesicular influx (of immature $A$ components) is assumed to decrease  with the compartment's content in $A$ components and to vanish for a fully matured compartment.
Therefore a fully mature compartment becomes isolated from the influx and exits the system as part of the out-flux while a new compartment is created {\em de novo}. This corresponds to the {\em cisternal maturation} mechanism of Golgi transport \cite{emr:2009}. Our model thus includes both {\em vesicular transport} and  {\em cisternal maturation}. The balance between these two mechanisms strongly relies on stochastic effects and is controlled by the ratio of maturation to vesicle budding rates and the steady-state organelle size.

The model is inspired by the regulatory role played by Rab GTPases on organelles dynamics \cite{schimmoller:1998,zerial:2001,stenmark:2001,stenmark:2009}, but is designed to be general and molecule-independent. The membrane of vesicles and organelles is discretised into patches of different membrane composition, to which  different biochemical identities can be assigned. The molecular identity of a membrane patch is defined by the presence of components that recruits proteins involved in membrane transport, but also by lipids that  influence the  membrane properties by changing  biophysical parameters such as the membrane bending rigidity. The maturation of membrane identity can involve the so-called Rab cascade, by which the activation of one Rab inactivates the preceding Rab along the pathway \cite{grosshans:2006}, which is  thought to operate both in endosomes \cite{rink:2005} and in the Golgi  \cite{rivera:2009,pfeffer:2010b}, but also the direct conversion of molecular components by enzymes, such as the glycosylation of proteins and lipids in the Golgi \cite{stanley:2011}.

\subsection*{Model description} We assume that the vesicles responsible for the influx and the outflux are of similar size, and we divide the membranes into patches of equal area so that a transport vesicle contains $N_{v}$ sites. A number $n_{v}$ of these sites are markers of the early ($A$) identity or the late ($B$) identity for incoming or outgoing vesicles, respectively, while the rest contains neutral species. The compartment state is then entirely defined by the total number of sites $N$ and the numbers $N_A$ and $N_B$ of sites containing $A$ and $B$ components. We define the relative fraction of $B$ components $\phi=N_{B}/(N_{A}+N_{B})$. In order to study the dynamics of the compartment, one must specify how the different kinetic processes depend on the compartment size and composition. We adopt the following notations for the influx of $A$ components, the maturation flux from $A$ to $B$, and the exiting flux of $B$ components:
\be
J_{A,in}=J(N, \phi)n_v\quad,\quad J_{A\rightarrow B}=k_m(\phi)N_A\quad,\quad J_{B,out}=K(\phi) n_v N_B
\label{fluxes}
\ee
In case the number of mature sites in the compartment satisfies $N_B<n_v$, the budding vesicle is assumed to remove all $N_B$ sites, and to contain a remaining number $N_v-N_B$ neutral sites, so that $ J_{B,out}=K(\phi)N_B^2$ (see Supporting Materials).
The different parameters involved in \eq{fluxes}  are defined below. The dynamics of the compartment is governed by the following set of stochastic transitions:
\begin{equation}
\begin{split}
\text{Fusion : } (N_{A},N_{B}) \rightarrow & (N_{A}+n_v,N_{B}) \text{ at rate }J(N, \phi)\\
\text{Maturation : } (N_{A},N_{B}) \rightarrow & (N_{A}-1,N_{B}+1) \text{ at rate }k_{m}(\phi)N_{A}  \\
\text{Budding : } (N_{A},N_{B}) \rightarrow & (N_{A},N_{B}-n_v)\text{ at rate }K(\phi)N_{B} \quad  {\rm if}\  N_B\geq n_v \\
\text{  } (N_{A},N_{B}) \rightarrow & (N_{A},0)\text{ at rate }K(\phi)N_{B} \quad  {\rm if}\  N_B<n_v \\
\end{split}
\label{dyneqs_stochastic}
\end{equation}

At the mean-field level the temporal evolution of the different components is given by:
\be
\dot N_A=J(N,\phi)n_v-k_m(\phi) N_A\qquad \dot N_B=k_m(\phi) N_A-K(\phi)n_vN_B
\label{dyneqs}
\ee
This expression is valid assuming that $N_B\geq n_v$ and that the ratio of active species $(N_A+N_B)/N$ in the compartment at $t=0$ matches the ratio $n_v/N_v$ in the transport vesicles. It is derived in the Supporting Materials, where the case $N_B<n_v$ is also discussed in more details.
This leads to the self-consistent equations for the mean-field steady-state
\be
\frac{N}{N_v}=\frac{J(N,\phi)}{(1-\phi)k_m(\phi)}\qquad  \phi=\frac{k_m(\phi)}{k_m(\phi)+K(\phi)n_v}
\label{s-s}
\ee

 We show below that the dynamics of the system can be separated into two main classes, regardless of the details of the functional form for the different rates.  The choices of functional form for the different fluxes discussed below is motivated by phenomenology rather than actual microscopic models or quantitative measurements. Other choices are possible, and will make quantitative differences. Our claim is that our general conclusions regarding the existence of these two classes are model independent.

\begin{itemize}
\item The flux of incoming vesicles $J(N, \phi)$ may depend on the size and composition of the receiving compartment. One can expect that $J$ is independent of the compartment size if it is limited by uni-dimensional diffusion ({\em e.g.} along a fixed number of cytoskeleton tracks) and  $J\sim \sqrt{N} $ (linear with the compartment size, and assuming a spherical compartment) if it is limited by three-dimensional diffusion. We require that $J$ increases sub-linearly with the compartment size, as  $J\sim N$ would not lead to a steady-state in our model. With regards to the membrane composition, homotypic fusion suggests that  $J$ decreases with increasing fraction $\phi$  of mature components. We assume here that no immature vesicles fuse with a fully mature compartment ($J(\phi=1)=0$), and study two different models: (i) a constant influx that abruptly vanishes when $\phi=1$, and a linear dependence: $J\propto(1-\phi)$, as simple choices.

\item Maturation ($J_{A\rightarrow B}$) is assumed to be a one step process with a rate $k_m$ that may involve some cooperativity and depend on the local concentration of $B$, hence of $\phi$. As an example, in a Rab cascade a Rab A recruits the effectors (GEF) which attracts another Rab B. Subsequently, the Rab B can recruit other effectors (GAP) which favor the unbinding of Rab A \cite{grosshans:2006}. These complex interactions can lead to a certain amount of cooperativity. This is taken into account here by writing $k_m(\phi)=k_m(1+\alpha\phi)$, where $k_m$ is the basal maturation rate and $\alpha$ represent the catalysing effect of neighbouring $B$ components for the maturation of $A$ components. A high value of $\alpha$ is a simple way to implement a switch-like behaviour between early and late identities, which could result from the feedback loops in the Rab cascade \cite{delconte:2008,binder:2012}

\item Vesicle budding ($J_{B, out}$) is assumed to extract specifically the $B$ components, and the budding rate $K(\phi)$ could be sensitive to the local composition of $B$ components, {\em e.g.} if several $B$ sites are needed to create a vesicle. After each budding event, $n_v$ sites of type $B$ are removed in a vesicle of size $N_v$. If the compartment contains a smaller number of mature sites $N_B<n_v$, all these sites are removed by the budding of one vesicle and both $N_B$ and $\phi$ vanish.
\end{itemize}

We have assumed that maturation and budding depend on local properties (the concentration), but not on the compartment size. This assumption could break down for small compartments, where maturation could be influenced by the total number of $B$ components in the compartment, and budding coud be reduced due to mechanical effects. We disregard these complexities here to reduce the number of parameters. The present model does of course not capture the full complexity of  real cellular organelles. In addition to coated vesicles, transport between cisternae within the Golgi apparatus might also proceed via of  membrane tubules connecting different cisternae \cite{marsh2004direct}. If such connections are transient, they can be described at a coarse-grained level within our framework of composition-dependent fluxes. Another important issue in Golgi dynamics is the recycling of Golgi resident enzymes. Recycling is essential in the cisternal maturation model to ensure that the enzymes remain at a particular location within the Golgi, and is expected to proceed via retrograde (trans-to-cis) vesicular transport \cite{Malhotra:2006}. Such complexity is not included in our model, which aims at answering well defined questions: what are the conditions under which a steady-state mixed compartment  undergo full maturation, and how relevant is this process to the net out-flux. Our analysis shows that the answer to these questions requires a stochastic analysis of compartment concentration fluctuations, which possess universal features that we identify. Including additional dynamical fields, such as the concentration of enzymes responsible for maturation, will be interesting, but is left for future work.

\subsection*{Output parameter}
The initial state is a compartment equivalent to one immature vesicles ($N=N_v,\ N_A=n_v,\ N_B=0$). The compartment evolves toward and fluctuates around a steady-state in which the vesicular outflux compensates the vesicular influx. However, this  ``steady-state'' has a finite life-time, if it is reached at all. The compartment will necessarily reach  full maturation ($\phi=1$) at some point due to stochastic fluctuations, and become isolated from the influx. In order to quantify the fraction of vesicular transport contributing to the total out-flux, we propose to compute the {\em output parameter} $\eta$ defined as follows:
\begin{equation}
\eta=\frac{\langle N_{ves} \rangle}{\langle N_{ves}\rangle+\langle N_{mat}\rangle}
\label{eta}
\end{equation}
where $\langle N_{ves}\rangle$ is the average number of matured - type B - vesicles emitted by the compartment before full maturation ($\phi=1$), and $\langle N_{mat}\rangle$ is the average size (measured in vesicle-equivalents) $N/N_v$ of the fully matured compartment. The out-flux is dominated by vesicular transport if $\eta \simeq 1$ and by full compartment maturation if  $\eta \ll 1$.

\section*{Results}

In this section, we first present analytical results for a simplified model where fusion, maturation and budding occur at constant rates and $n_v=1$. We then explore the case of composition-dependent rates numerically and explain the main difference that arise based on the mean-field dynamics of the system. The effect of size-dependent fusion, which we found to be minor, is discussed in the Supporting Material.

\subsection*{Analytical solution for constant rates}
We start by analysing the simplest version of the model with $n_v=1$ and where the influx $J$, the maturation rate $k_m$ and the fusion rate $K$ are all constant, independent of the compartment size and composition. In this case, one may obtain an analytical solution for the isolation time of a compartment, and an approximate analytic expression for the output parameter $\eta$. The mean first passage time to isolation, namely the time needed to reach $N_A=0$ starting at $N_A=1$, can be calculated exactly, as the dynamics of $A$ components in the compartment is independent of the dynamics of the $B$ components. The transition rates governing the evolution of $N_{A}(t)$ are:
\be
N_{A}\xrightarrow[]{J} N_{A}+1 \qquad N_{A}\xrightarrow[]{k_{m}N_{A}} N_{A}-1
\label{rate_simple}
\ee

The average time $\tau_n$ needed to reach $N_{A}=0$ starting from a state $N_{A}=n$ following this simple stochastic process satisfies the classical recursion law for mean first passage times \cite{vankampen:2007}, which is derived in the Supporting Material. For the rates defined in \eq{rate_simple}, the recursion relation reads:
\begin{equation}
-1=k_{m}n(\tau_{n-1}-\tau_n)+J(\tau_{n+1}-\tau_n)\label{recur_tau}
\end{equation}

An expression for the average time to full maturation starting from a newly created compartment, $\tau_1$, can be obtained by solving \eq{recur_tau} recursively to obtain the expression:
\be
\frac{\tau_{n+1}-\tau_n}{n!}=\left(\frac{k_m}{J}\right)^n\tau_1-\frac{1}{J}\sum_{i=0}^{n-1}\frac{1}{(n-i)!}\left(\frac{k_m}{J}\right)^i
\ee
For $n \gg J/k_{m}$, the mean field analysis, \eq{dyneqs}, suggests that the system first evolves toward the (quasi) stationary state given by \eq{s-s}, in a typical time of the order of $1/k_{m}$ independent of $n$, and remains there for a (potentially long) time before full maturation. Therefore we expect that $\lim_{n\gg J/k_m}(\tau_{n+1}-\tau_n)/n!=0$, which leads to an explicit formula for $\tau_{1}$:
 \begin{equation}
\tau_1=\frac{e^{\frac{J}{k_{m}}}-1}{J}=\frac{e^{N(1-\phi)}-1}{k_{m}N(1-\phi)}
\label{tau1}
\end{equation}
where $N$ and $\phi$ are the steady-state average size and composition of the compartment (\eq{s-s}).

To estimate the value of the output parameter $\eta$ (\eq{eta}), we must compute the average number of vesicles emitted before compartment isolation $\langle N_{ves}\rangle$, and the average size of the fully matured compartment $\langle N_{mat}\rangle$. Calculating these quantities analytically is difficult, because it requires solving the two-dimensional isolation problem for $N_A$ and $N_B$. An estimate of the size of the matured compartment is $\langle N_{mat}\rangle\simeq N\phi$. This amounts to saying that isolation occurs due to a temporary lack of incoming vesicles, while the number of $B$ components retains its steady-state value due to a balance between maturation and vesicle secretion. An estimate of the amount of emitted vesicles before maturation can be obtained by supposing that the system spends most of its time undergoing small fluctuations around the steady-state, so that: $N_{ves}\approx KN\phi\tau_1=e^{N(1-\phi)}-1$. The fact that newly formed compartments are initially small and may reach full maturation without ever reaching the steady-state modifies this estimate. This can be crudely taken into account by considering that the initial compartment made of one vesicle can mature directly and become isolated in only one step. This one-step maturation event corresponds to $N_{ves}=0$ and $N_{mat}=1$ and happens with the probability:
\begin{equation}
p_{1}=\frac{k_{m}}{k_{m}+J}
\label{p1}
\end{equation} 
Taking this into account, we obtain the following estimates:
\begin{equation}
\begin{split}
N_{ves} \approx & (1-p_{1})(e^{N(1-\phi)}-1) \\
N_{mat} \approx & p_{1}+(1-p_{1})N\phi
\end{split}
\label{vesmat}
\end{equation}
from which we get the approximate results:
\be
\frac{\eta}{1-\eta}=\frac{\frac{J}{k_m}\left(e^{\frac{J}{k_m}}-1\right)}{1+\frac{J^2}{Kk_m}}=\frac{N(1-\phi)\left(e^{N(1-\phi)}-1\right)}{1+N^2\phi(1-\phi)}
\label{eta2}
\ee

\subsection*{Numerical simulation under constant maturation and budding rates} We performed numerical simulations of the stochastic dynamics  of \eq{dyneqs_stochastic}, following the Gillespie scheme described in the Supporting Material. We first restrict ourselves to the case where the  maturation and fusion rates are constant and where the content of transport vesicles corresponds to a single membrane patch: ($N_v=n_v=1$).  We focus our analysis on two experimentally observable quantities: the dynamical features of individual time traces of size and concentration fluctuations of compartments, and the size distribution of fully mature compartments. Averaging over many simulations, we construct a phase diagram for the output parameter as a function of the different exchange rates which illustrate the regions where maturation and vesicular exchange dominate the compartment dynamics.

\subsubsection*{Constant influx}
\begin{figure}[t]
\centering
\includegraphics[width=6.75in]{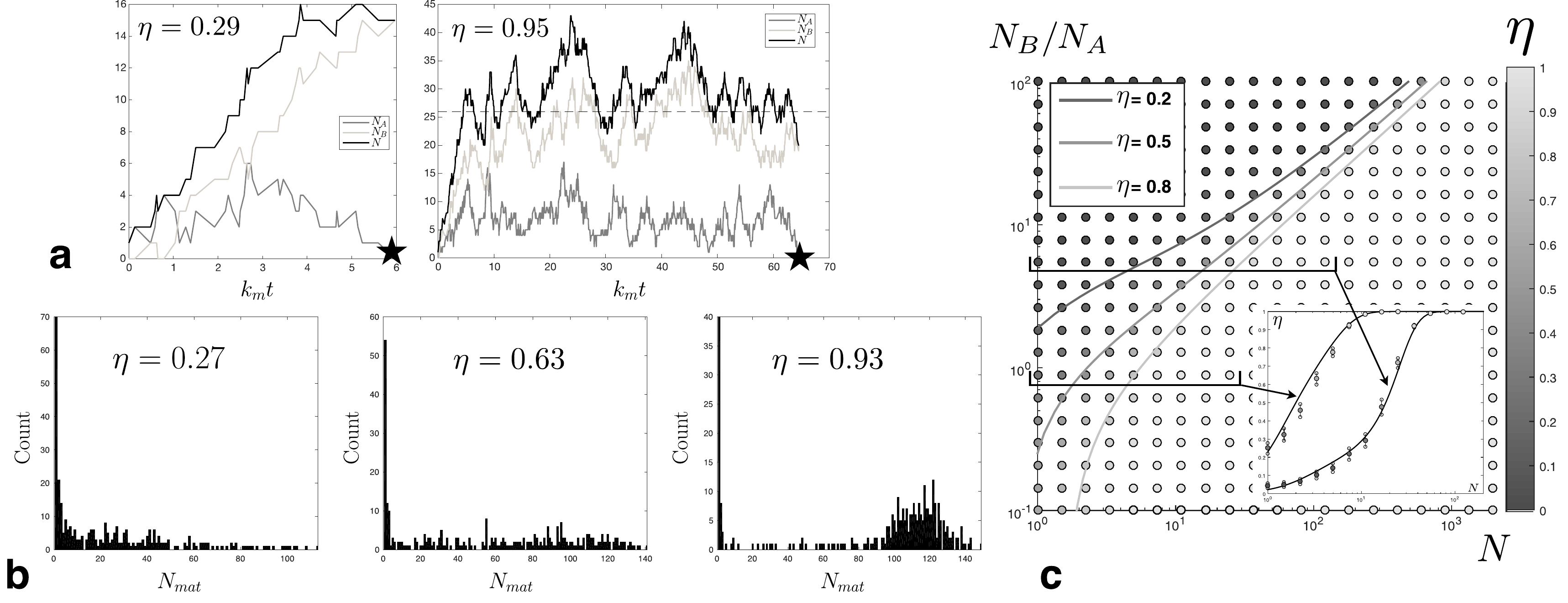}
\caption{\label{fig:Jconst} Simulation results for constant rates. {\bf a.} Typical time trace showing the fluctuations of the compartment's content in $A$ and $B$ sites ($N_A$ and $N_B$) and total size $N/N_v=N_A+N_B$ as a function of the dimensionless time $k_mt$,  in the maturation ($\eta=0.29$, $J/k_m=4.16$ and  $K/k_m=0.1905$) and vesicular  ($\eta=0.95$, $J/k_m=6$ and  $K/k_m=0.3$) regimes. The steady-state value of $N$ is shown as a dashed line and full maturation is indicated by a black star. {\bf b.} Size distribution of fully matured compartments ($N_{mat}$) obtained from 320 independent simulations (for three different values of $\eta$). The steady-state parameters are $N=121$  $N_B/N_A= 33.6,\ 23.4,\ 16.2$ from left to right).
{\bf c.} Phase diagram for the value for the output parameter $\eta$ as a function of the pseudo steady-state compartment size $N$ and ratio $N_B/N_A$, showing the transition between the {\em vesicular exchange} ($\eta\simeq1$) and {\em compartment maturation} ($\eta\simeq0$) regimes. The three gray lines represent constant values of $\eta$ as given by the approximate analytical computation of \eq{eta2}. (Inset) Cuts through the phase diagram varying the compartment size for two fixed pseudo steady-state compositions. The black lines are obtained using \eq{eta2} and the dots are the simulation results and the associated error bars. }
\end{figure}

We first report the results of simulation when all the exchange rates (Influx $J$, maturation rate $k_m$ and budding rate $K$) are constant, as in the analytical calculation of the previous section.
A typical evolution of the number of $A$ components and the total number of components in the compartment is shown in \fig{fig:Jconst}.a. Depending on the ratio of maturation to budding rate, the system either displays strong fluctuations around the steady-state (\eq{s-s}), which eventually lead to the complete maturation of all $A$ components, or exhibits full maturation before reaching the steady-state. 
 
 After many independent realisations of the maturation process, one obtains a distribution of values for the size of the fully mature compartment $N_{mat}$ and the number of vesicles exported prior to the full maturation $N_{ves}$, from which the output parameter $\eta$ (\eq{eta}) can be calculated. \fig{fig:Jconst}b shows that the distribution of $N_{mat}$ is rather broad in the maturation dominated regime (small values of $\eta$) and shows a peak at the steady-state compartment size in the regime dominated by vesicular exchange ($\eta\simeq 1$). The values of $N_{mat}$ and $N_{ves}$ for different parameters are represented as scatter plots in the Supporting Material. They are well fitted by assuming that the compartment follows the mean field dynamics given by \eq{dyneqs} and assuming that full maturation occurs after a time $t_{mat}$:
\begin{equation}
N_{ves}=\int_{0}^{t_{mat}}KN_B(t)dt \qquad N_{mat}=N_{B}(t_{mat})\label{eq:scatter}
\end{equation}
This shows that, for the parameters of \fig{fig:Jconst}, the main source of fluctuation comes from the distribution of the isolation time $t_{mat}$.

 The phase diagram for the output parameter is shown in \fig{fig:Jconst}c as a function of the steady-state compartment size and distribution. The same diagram is shown as a function of the model parameters $J/K$ and $k_m/K$ in the Supporting Material. The two extreme mechanisms of {\em vesicular exchange} ($\eta\approx 1$) and {\em compartment maturation} ($\eta\approx 0$) are observed for extreme values of the parameters, namely large steady-state compartment size $N$ for the former and high value of the steady-state fraction $\phi$ for the later. However, the phase diagram shows a  richer picture, with a  gradual transition between the two mechanisms upon variation of the ratio of maturation to budding rates for intermediate compartment size. The analytical calculation of the output parameter (\eq{eta2}) faithfully reproduces the numerical results, except for very small compartments.
 
 The size distribution of mature compartments (\fig{fig:Jconst}b) shows a large peak at very small size ($N_{mat}\simeq 1$). This corresponds to cases where a young compartment undergoes direct maturation prior to any, or after a few, fusion and budding events. Such fully mature compartment is structurally indistinguishable from a budded vesicle, and it may seem arbitrary to include the former in the maturation flux and the latter in the vesicular flux, as is done in the definition of the output parameter (\eq{eta}). This is nevertheless reasonable within our model, where the difference between the two fluxes is a matter of kinetic processes rather than a difference of structure. For completeness, we show in the Supporting Material an equivalent of \fig{fig:Jconst}c for two alternative definitions of the output parameter, either removing all direct maturation events, or removing all event where the fully mature compartment is of unit size ({\em e.g.} following the same number of fusion and budding events). The difference is only quantitative, and only appreciable for very small steady-state size.
  
\subsubsection*{Composition-dependent influx}
\begin{figure}[t]
\centering
\includegraphics[width=6.75in]{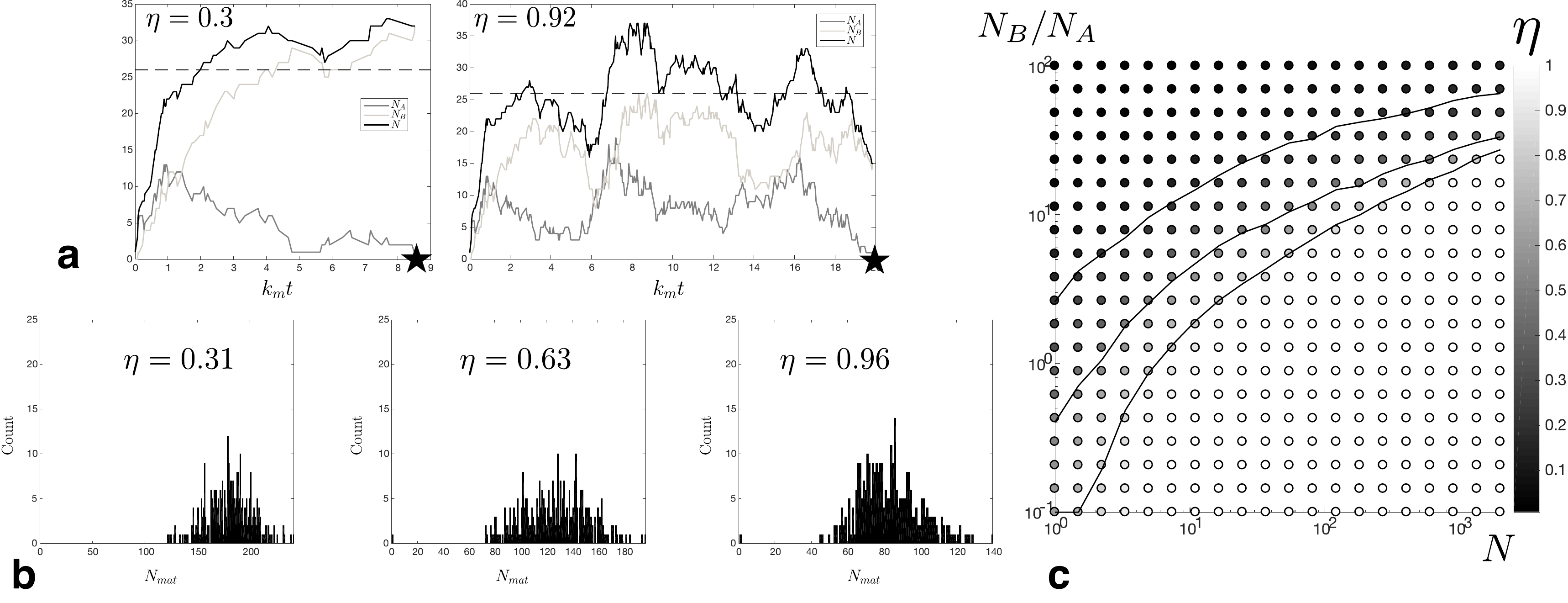}
\caption{\label{fig:Jvar} Simulation results with homotypic fusion.  Identical to \fig{fig:Jconst}, but for a concentration-dependent influx $J=J_0(1-\phi)$. {\bf a.} Typical time traces of the compartment composition for two values of the output parameter ($\eta= 0.3$ corresponds to $J/k_m=26$ and  $K/k_m=0.06$, and $\eta= 0.92$ corresponds to $J/k_m=26$ and  $K/k_m=0.5$). {\bf b.} Size distribution of fully matured compartments (the steady-state parameters are $N=121$ and  $N_B/N_A=23.4,\ 11.3,\ 5.46$ from left to right). {\bf c.} Phase diagram for the value of the output parameter $\eta$ as a function of the pseudo steady-state compartment size $N$ and ratio $N_B/N_A$ of the two components. The three gray lines represent constant values of $\eta=0.2,\ 0.5$ and $0.8$ as in \fig{fig:Jconst}.}
\end{figure}

The previous model with a composition-independent influx possesses the rather arbitrary feature that the influx abruptly drops to zero when the compartment reaches full maturation. While it is possible that homotypic fusion could permit a steady influx of immature vesicles with only one, or a few, immature site in the compartment, one may also expect a more gradual dependence of the influx with the compartment concentration. We present in \fig{fig:Jvar} the same results as in \fig{fig:Jconst}, but with an influx that linearly decreases with the composition of the compartment: $J=J_0(1-\phi)$ (henceforth called the homotypic fusion model). The general features of the phase diagram are conserved, namely the dominance of maturation for small compartments and large steady-state fraction of $B$ components. A composition-dependent influx however brings three important qualitative differences: {\em(i)} the vesicular exchange-to-maturation transition shows a much weaker dependence upon the steady-state compartment size (for large sizes), {\em (ii)} the size distribution of fully mature compartment is always peaked around a large size and direct maturation of incoming vesicles ($N_{mat}=1$) is very rare, and finally {\em (iii)} the time trace of the compartment composition in $A$ and $B$ species seem anti-correlated, and display oscillations for large $\eta$. We show below that these features can be understood by analysing the mean-field dynamics of the system.

\begin{figure}[t]
\centering
\includegraphics[width=6.75in]{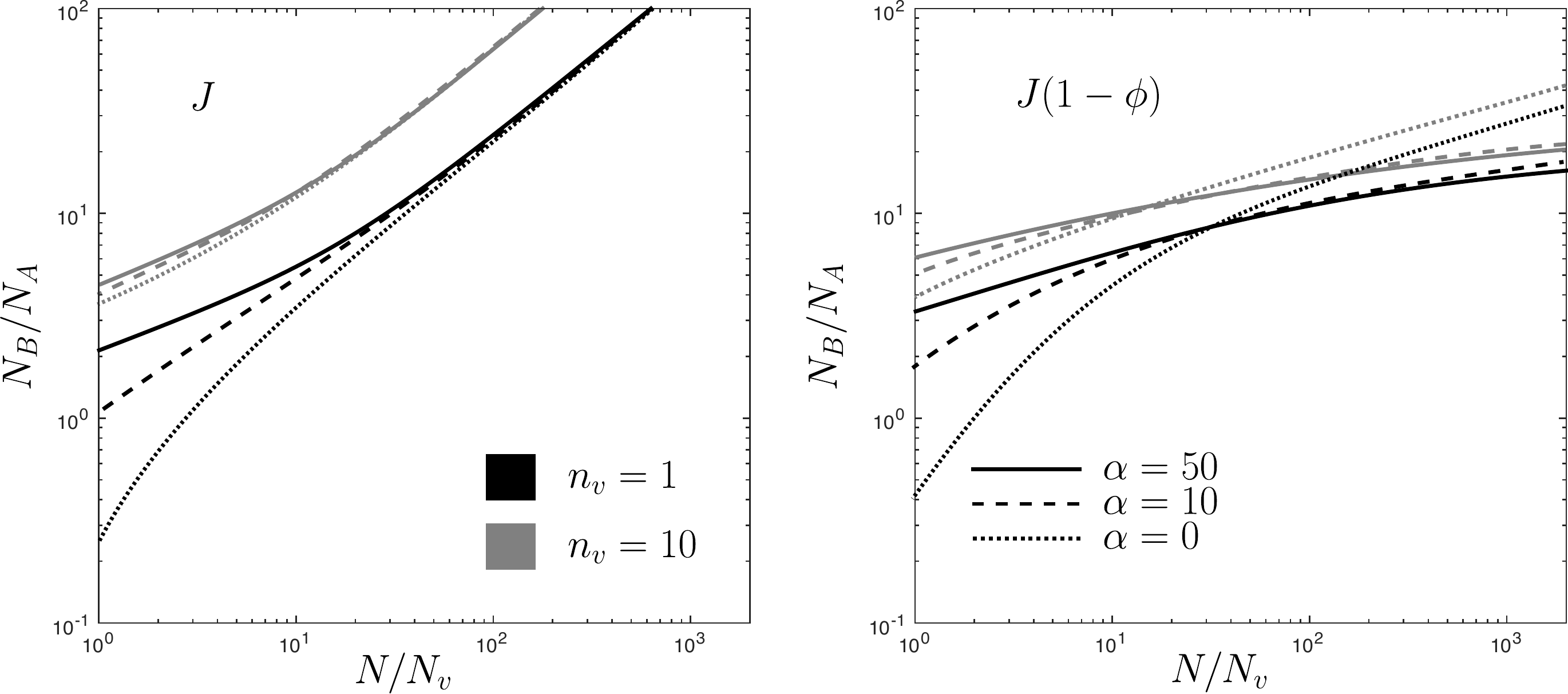}
\caption{\label{fig:refine} Level curves ($\eta=0.5$)  extracted from phase diagrams similar to the one of \fig{fig:Jvar} for composition-dependent rates of the different kinetic processes in \eq{fluxes}. The vesicular influx $J(N,\phi)$ is either constant (left panel) or controlled by linear homotypic fusion ($J=J_0(1-\phi)$ - right panel). The effect of cooperativity in the maturation process is seen by varying the  coefficient $\alpha$ (with a maturation rate $k_m(\phi)=k_m(1+\alpha\phi)$). The number of components per vesicle is $n_v=1$ (black curves) and $n_v=10$ (grey curves). The curves corresponding to  the phase diagram of Figs.\ref{fig:Jconst} and \ref{fig:Jvar} are the dotted black line in the left and right panels, respectively.}
\end{figure}

\subsubsection*{Effect of the other parameters}
We now relax the assumption  that the entire content of incoming vesicles mature as one entity and assume that the incoming and outgoing vesicles contain $N_v$ independent membrane sites, of which $n_v$ are active sites as in \fig{fig:sketch}b. We also include the possibility for cooperative maturation ($k_m(\phi)=k_m(1+\alpha\phi)$ in \eq{fluxes}). The impact of these  parameters on the transition between \textit{compartment maturation} and \textit{vesicular exchange} is shown on \fig{fig:refine}, which displays the boundary corresponding to $\eta=0.5$ for constant and composition-dependent influx,  varying the maturation cooperativity parameter $\alpha$ and using two values of the number of active molecules in the incoming vesicles: $n_v=1$ and  $n_v=10$. The latter value is  the typical  number of Rab molecules in a transport vesicle \cite{takamori:2006}. The global trends are as follows: cooperativity in the maturation process disfavours maturation for very small compartments but favours it for large  compartments ($N\gtrsim 10 N_v$). The number of  maturing components per vesicle $n_v$ has a strong impact on the position of the boundary - disfavouring full compartment maturation - for constant influx, but only has a weaker effect with composition-dependent influx. Combining homotypic fusion with high maturation cooperativity renders the transition almost independent on the compartment size. The full phase diagrams, displayed in the Supporting Material, show that the width of the transition does not depend on $n_v$ or $\alpha$, but only on whether the influx is composition-dependent.

As a final refinement of the model, we also consider the possibility that  the maturation of $A$ into $B$ components might involve several steps. This is for instance the case of the 
Rab cascade which includes the recruitment of GEF and GAP molecules. At our level of description,  the
principal consequence of having several reaction steps is that the maturation waiting time might not be exponentially distributed. In the Supporting Material,
we study the extreme case where  the maturation of a single $A$ component involves a very large number of independent steps and becomes an almost deterministic process with a fixed waiting time. Remarkably, we show that the average maturation time $\tau_1$ is still given by \eq{tau1}. This suggests that our results 
do not depend much on the distribution of maturation times.

\subsection*{Role of homotypic fusion}
The peculiar dynamics of compartments experiencing composition-dependent influx, namely the anti-correlated nature of composition variations of  $A$ and $B$ components prior to maturation,  the peaked size distribution of fully matured compartment, and the weak dependence of the full maturation probability on the compartment steady-state size, can all be understood from the mean-field dynamics given by \eq{dyneqs}.
The steady-state  (\eq{s-s}) is always stable in our model, but the relaxation toward the fixed point is either overdamped if $J$ is constant, or underdamped (oscillatory) with homotypic fusion ($J\sim(1-\phi)$), beyond a critical ratio of $\bar k_m=k_m/K$. This is analysed in the Supporting Material. In the absence of cooperativity in the maturation process ($\alpha=0$), the critical ratio obeys
\be
4\bar k_m^3+4\bar k_m^2-1=0\quad{\rm or}\quad k_m/K\simeq0.42.
\ee
This threshold decreases with increasing $\alpha$ (see supporting Material).

Far above threshold ($k_m/K\gg 1$), the compartment first grows through a burst of vesicle injection $(N_A \gg 1,N_B\approx 0)$, followed by a slower maturation process $(N_A \approx 0,N_B \gg 1)$. During the latter, only a few stochastic maturation steps are needed to reach full maturation. The likelihood of this to happen becomes weakly dependent on the compartment size, and on the level of coarse-graining $n_v$. 
The size distribution of fully mature compartments reflects this dynamics. If $J$ is constant, it is peaked at the mean-field steady-state size (\eq{s-s}) when  vesicular exchange dominates the out-flux ($\eta \lesssim 1$), but is rather broad in the maturation dominated regime, see \fig{fig:Jconst}b. With composition dependent influx, full maturation occurs predominantly during phases of the spiralling trajectories where $N_A$ is small, leading to large matured compartment with a well-defined  range of sizes \fig{fig:Jvar}b. Finally, the absence of direct maturation of incoming vesicles ($N_{mat}=1$) with homotypic fusion can be understood as follows: A given steady-state $(N,\phi)$ corresponds to a smaller value of the ratio of influx over maturation rate for a constant influx ($J_0/k_m=N(1-\phi)$) than for a composition-dependent influx ($J_0/k_m=N$). The likelihood of direct maturation of incoming vesicles (the probability $p_1$ in \eq{p1}) prior to a fusion event is thus strongly reduced by homotypic fusion.

The linear stability analysis of the mean-field equations can easily be extended to any functional form one wishes to explore, but what really matters is whether the system is in the over-damped or underdamped regim. As we show above, this is mostly controlled by homotypic fusion. Using a linear stability analysis method, one can easily convinced oneself that if the budding rate increases with the concentration of $B$ component, as could be expected if budding is a cooperative process, the range of parameter for which the dynamics is over-damped is increased (the fixed point is stabilised) and  full maturation is disfavoured. If the budding rate is assumed to follow a Michaelis-Mentens kinetics with saturation at high concentration, the effect is reversed, and full maturation is favoured.

\section*{Discussion}

Are membrane-bound organelles along the cellular secretory and endocytic pathways steady-state structures receiving, processing, and exporting transiting cargoes, or are they transient structures that grow by fusion of smaller structures until they reach full maturation? To address this question from a theoretical viewpoint, we have developed a minimal stochastic model combining compartment maturation and vesicular exchange. The model is based on two fundamental assumptions: that there exists a steady-state where the outflux balances the influx, and that there exists a particular composition of the system for which it becomes committed to full maturation. The former assumption is a wide spread concept for cellular organelle in general \cite{chan:2012}, and the endomembrane system in particular \cite{vanmeer:2008}. The latter is supported by observations of early endosomes \cite{rink:2005} and \textit{S. cerevisiae} Golgi cisternae  \cite{Matsuura:2006,losev:2006}. We show that the organelle's dynamics is essentially controlled by two parameters:  the ratio of vesicle injection to budding rate which controls the steady-state size of the organelle and the ratio of biochemical maturation to budding rate, which controls the organelle's steady-state composition.

Our main results are summarised Figs.\ref{fig:Jconst},\ref{fig:Jvar}. For low maturation rates (compared to the budding rate), the  flux of mature material leaving the system is predominantly composed of budded  vesicles of mature components. The organelle remains at steady-state for a long time while exporting vesicles of mature component. When full maturation occurs, the mature compartment has a size distribution peaked around the steady-state size. For high maturation rates on the other hand,  full maturation typically occurs before the steady-state is reached and the outflux is mostly composed of fully mature compartment. Importantly, full maturation can follow one of two types of dynamics: the trajectory in the phase space (size and composition) is either a quasi random walk eventually hitting the full maturation threshold, or a spiralling trajectory around the fixed point that reaches full maturation in a quasi deterministic fashion. The difference is important, as the  fully mature organelles produced by the former mechanism are unlikely to be very large and have a broad size distribution, while those produced by the latter can be large and have a peaked size distribution. A necessary condition for the occurrence of deterministic maturation is that the influx decreases as the organelle becomes more mature, which is an expected consequence of homotypic fusion. Positive feedback in the maturation process renders full compartment maturation more deterministic.

Figs.\ref{fig:Jconst},\ref{fig:Jvar} suggest that the dominant export mechanism of an organelle could be inferred by analysing individual time series of the organelle's size and composition. Maturation of \textit{S. cerevisiae} Golgi cisternae from a {\em cis} to a {\em trans} identity appears fairly deterministic, with a gradual evolution from one identity to the other \cite{Matsuura:2006,losev:2006}. This suggests that this organelle is in the maturation-dominated regime helped by homotypic fusion, a conclusion reinforced by the observation that different time traces of compartment composition display reproductible dynamics  \cite{losev:2006}. Maturation of early endosomes into late endosomes appear much more stochastic, with large composition fluctuations \cite{rink:2005}. In this case our results suggest that vesicular export of the late membrane identity (which has not been investigated in \cite{rink:2005}) might dominate the dynamics. Note that the individual fuorescence tracks shown in \cite{rink:2005} seem to display large amplitude oscillations prior to full maturation, similar to the theoretical tracks shown in \fig{fig:Jvar}. In our model, these oscillations are a consequence of the spiralling (overdamped) trajectories in the phase space, a signature of homotypic fusion. This suggests that, although in the vesicular exchange regime, these endosomes are fairly close to the maturation transition, and that endosomal dynamics could be converted to a maturation dominated regime by small change of parameters, such as an increase of the ratio of maturation rate to budding rate. These considerations are clearly very preliminary, and a more systematic analysis is needed to reach a more definite conclusion, but these examples illustrate the intimate link between the temporal fluctuations of individual components and the time-average export dynamics of organelles.

\subsection*{Vesicular exchange and maturation in the Golgi apparatus.}
Are Golgi cisternae stable structures that receive and export material while retaining their identity, or are they transient structures that progressively mature from the {\em cis} to the {\em trans} identities? While evidences for the latter dynamics exist in endosomes and the Golgi cisternae of the yeast  \textit{S. cerevisiae}, this issue is not resolved for the stacked Golgi of most higher eukaryotes including mammalian cells, for which a long-lasting controversy exist between the {\em cisternal maturation} and {\em vesicular exchange} transport models. The reality might lie between these two extreme scenarios, and our model can in principle give a quantitative answer as to the fraction contributed by each mechanism to the total outflux, provided that the rates of vesicular influx, maturation and budding are known.
The challenge in comparing our theory to experiments lies in obtaining accurate values for these rates.
In the following, we concentrate on the dynamics of the Golgi apparatus, and we adopt the estimate $10^{-3}\leq K\leq 10^{-2}\unit{s^{-1}}$  for the budding rate for COPI vesicles \cite{wang:2008}. Values for the influx depend on whether one considers the Golgi ribbon of vertebrate cells which receives influx from the entire Endoplasmic Reticulum (ER)\cite{wei:2010}, for which we estimate $J\sim10-10^2\unit{s^{-1}}$, or individual Golgi mini-stacks formed at ER exit sites, for which we estimate $J\sim10^{-2}-10^{-1}\unit{s^{-1}}$ \cite{thor:2009,warren:2013}. We also take the latter value for the influx toward individual cisternae of the dispersed Golgi of yeast \textit{S. Cerevisiae}.  Maturation of Golgi cisternae of the yeast \textit{S. Cerevisiae} has been monitored by live imaging, yielding an isolation time of order  $\tau\approx 10^2\unit{s}$ \cite{Matsuura:2006,losev:2006,bhave:2014}. 
For a stacked Golgi, and assuming  the isolation time is of order the typical transit time of cargoes across the stack (a lower bound corresponding to the {\em cisternal maturation} model), it is of order $\tau\approx 10^3\unit{s}$ \cite{mironov:2001,patterson:2008}. With these  parameters and assuming a constant influx, the maturation rate can be obtained from \eq{tau1} and the steady-state size and composition of cisternae from \eq{s-s}. We find $k_m\sim 10^{-2}-10^{-1}\unit{s^{-1}}$ for  \textit{S. Cerevisiae}, and $k_m\sim 10^{-3}-10^{-2}\unit{s^{-1}}$ for Golgi ministacks. The output parameter (\eq{eta}) is of order $\eta=0.1-0.5$ for \textit{S. Cerevisiae} and $\eta=0.5-0.9$ for ministacks, placing the former in the {\em maturation}  dominated regime (in agreement with previous studies \cite{losev:2006, papanikou:2009}) and the latter in the {\em vesicular exchange} dominated regime (as studies based on cargo transport dynamics also concluded \cite{patterson:2008,dmitrieff:2013a}). The predicted steady-state size $N\sim 10-100$ for both  is reasonable. 
These conclusions remain qualitatively valid with the homotypic fusion model and a composition-dependent influx.
 It is unclear whether the rather simple model developed here is adequate to describe the Golgi ribbon, itself a compact assembly of somewhat interconnected Golgi mini-stacks \cite{wei:2010}. With the corresponding parameters, we find a maturation rate $k_m\sim 1-10\unit{s^{-1}}$, corresponding to $\eta\sim 0.5-0.9$, also in the {\em vesicular exchange} dominated regime, and a steady-state size $N=10^4-10^5$. 

It is interesting to notice that the Golgi in the two different cell types are predicted to be on opposite sides of the boundary between {\em vesicular exchange} and {\em cisternal maturation}, and  also have very distinct morphologies (dispersed vs. stacked), suggesting a possible correlation between dynamics and morphology. When \textit{S. Cerevisiae} is starved in a glucose-free environment, the isolation time of Golgi cisternae increases to $\tau = 3\times 10^2\unit{s}$ \cite{levi:2010}. Within our model, an increase of $\tau$ can result from an increase of $J$ or a decrease of $k_m$ (\eq{tau1}). The latter seems much more likely than the former  in a starved situation. The slowing down of Golgi kinetics leads to a shift toward the {\em vesicular exchange} dominated regime ($\eta\sim 0.2-0.7$). Remarkably, the Golgi structure is also modified, and resembles the stacked Golgi structure in \textit{P. Pastoris} \cite{levi:2010}.  This observation strengthens the proposal that correlations exist between Golgi structure and transport kinetics. The Golgi of  \textit{P. Pastoris} could be intermediate, both in terms of transport dynamics and structure, between the Golgi of \textit{S. Cerevisiae} and the stacked Golgi of most eukaryotes, as it is at the same time stacked and shows continuous cisternae turnover akin to {\em cisternal maturation} at its {\em trans} face \cite{mogelsvang:2003}. Unfortunately, we could not evaluate the output parameter for \textit{P. Pastoris}, as there is to our knowledge no available quantitative data of Golgi transport kinetics in this organism.

\subsection*{Quantitatively testable predictions of the model}
{\bf Statistics of individual time traces.} Our model makes a number of experimentally testable predictions, regarding the relationship between the dynamics of individual compartments, and in particular the concentration fluctuations over time, the size distribution of mature compartment, and the dominant export mechanism. The time average dynamics of the system can in principle be obtained form a statistically significant set of individual time series of compartment size and composition. The crude rule of thumb is that if an organelle spends a significant fraction of  time at a quasi-steady state, possibly undergoing large fluctuations around it, its dynamics is likely to be in the vesicular exchange regime. The distribution of maturation time, experimentally more accessible than the size distribution of mature compartments as those might undergo further homotypic fusion, is less discriminatory as it is model dependent. While a broad distribution is a signature of a maturation-dominated regime with a constant (or weakly composition-dependent) influx, a peaked distribution could correspond to a regime dominated by vesicular exchange, but also to a maturation-dominated regime with a composition-dependent influx (related to homotypic fusion), which renders full maturation almost deterministic. Thus the combined analysis of individual time trace and the distribution of isolation time can inform us both on the dominant export mechanism, and on the specificity of the fusion, maturation, and budding processes.

\vspace{2mm}\noindent 
{\bf Rate of cargo transport.}
Our model produces an interesting prediction with regards to cargo transport. At steady-state, a cargo exported in vesicles should leave organelles such as the Golgi at a rate $J$ while a cargo unable to enter transport vesicle and relying solely on cisternal maturation should be exported at a (potentially much) slower rate $1/\tau_1\sim Je^{-J/k_m}$ according to the constant flux model, \eq{tau1}. This could apply to large protein complexes such as procollagen, whose progression through the Golgi stack, which is about twice as slow as that of smaller membrane proteins such as VSVG \cite{dmitrieff:2013a},  has 
been taken as evidence for cisternal maturation \cite{bonfanti:1998,mironov:2001}. Regulation of the export mechanism could be very important for the transport of such large cargoes that do not fit inside export vesicles. Our results suggest a possible mechanism for this regulation. The presence of such large cargo  as procollagen in Golgi cisternae could reduce the rate of vesicle secretion $K$, {\em e.g.} by mechanical means through an increase of membrane tension imposed by the distension of the cisternal membrane. This would favour full cisternal maturation and permits the progression of the large cargo through the Golgi stack. Interestingly, VSVG has been observed to move synchronously with procollagen when both are present in the Golgi  \cite{mironov:2001}, lending support to this regulatory mechanism. Note however that quantitative analysis of intra-Golgi transport suggests that procollagen transport does not solely rely of cisternal maturation \cite{dmitrieff:2013a}.

\section*{Conclusion}
The highly dynamical nature of intracellular organisation requires the exchange processes between organelles to be tightly regulated in order to yield robust directional flow of material through the cell. While the cell may to some extent be viewed as the steady-state of a complex dynamical system, specific budding, fusion and maturation events, which  shape its organisation, are inherently stochastic processes. Owing to the relatively small size of many cellular organelles, stochastic fluctuations must be accounted for in models of their dynamics. We have developed a stochastic dynamical model to study the interplay between maturation and exchange in intracellular trafficking. Our model can reproduce both the strong fluctuations of size and composition seen in early endosomes \cite{rink:2005} (\fig{fig:Jconst}) and the more deterministic maturation of individual Golgi cisternae in \textit{S. Cerevisiae}. It  includes as asymptotic limits the two extreme exchange mechanisms at the heart of the Golgi transport controversy \cite{emr:2009}; {\em vesicular exchange} and {\em compartment maturation}. We identify full compartment maturation as a first passage process, whose likelihood decreases with increasing organelle size. We also found that the interplay of homotypic fusion and cooperative maturation increases the probability of  full maturation and reduces its size dependence. These mechanisms therefore act as regulators to provide robustness to full compartment maturation against stochastic fluctuations.

\section*{Acknowledgments}
This work has received support under the program ``Investissements d'Avenir" launched by the French Government and implemented by ANR with the references ANR-10-LABX-0038.

\section*{Author contributions}
Q.V and P.S designed the research. Q.V performed the research. Q.V and P.S analyzed the data and wrote the paper.

\section*{Supporting citations}
Reference \cite{gillespie:1977} appears in the Supporting Material.

\bibliographystyle{ieeetr}

\end{document}


\title{Stochastic Model of Maturation and Vesicular Exchange in Cellular Organelles\\ \vspace{0.5cm} {Supporting Material}}
\date{\today}
\author{Quentin Vagne and Pierre Sens}
\maketitle

In this Supporting Material, we derive the recursion relation for the mean first passage time to complete compartment maturation, we show how to obtain the mean-field approximation of the stochastic compartment dynamics, we give details of the numerical procedure used for the simulations, and we provide additional simulation results.

\section{Recursion relation}
We start by deriving the classical equation on mean first passage times for a discrete one dimensional system submitted to a stochastic dynamics governed by transition rates. We consider a system described by a discrete variable $n \in \mathbb{N}$. The dynamics of $n$ is stochastic and the transition rate from the configuration $n$ to another configuration $m$ is written $R_{n\rightarrow m}$. We are interested in the quantity $\tau_{n}$ which is the average time needed to reach the configuration $0$ starting from configuration $n$. We call $p(n,t)dt$ the probability to reach $0$ between $t$ and $t+dt$ starting from $n$ at $t=0$. One can write an equation on the functions $p(n,t)$ by computing the probability to reach $0$ at time $t+dt$ starting from $n$. In order to do this, during a time $dt$ the system will either jump towards another configuration $m$ or stay in $n$ and then use a time $t$ to reach the configuration $0$. Mathematically written, it reads:
\begin{equation}
p(n,t+dt)=\sum_{m\ne n}R_{n\rightarrow m}dt\ p(m,t)+\left(1-\sum_{m\ne n}R_{n\rightarrow m}dt\right)p(n,t)
\end{equation}
This leads to a differential equation on $p(n,t)$: 
\begin{equation}
\frac{\partial p(n)}{\partial t}=\sum_{m\ne n}R_{n\rightarrow m}(p(m,t)-p(n,t))
\end{equation}
Multiplying by $t$ and integrating on the time leads to an equation for the mean first passage time $\tau_{n}$ to reach $n=0$ starting at $n$, which a classical result obtained in \cite{vankampen:2007} p298-303 in the restricted case of one-step processes:
\begin{equation}
\forall n \in \mathbb{N}\text{ : }-1=\sum_{m \ne n}R_{n\rightarrow m}(\tau_{m}-\tau_{n}).
\label{generic_recursion_tau}
\end{equation}
This recursion relation is used is the main text (Eq.(9)) to compute the mean first passage time to full maturation, Eq.(11).

\section{Mean-field approximation of the compartment dynamics}
The (inherently stochastic) compartment dynamics is defined by Eq.2 from the main text:
\begin{equation}
\begin{split}
\text{Fusion : } (N_{A},N_{B}) \rightarrow & (N_{A}+n_v,N_{B}) \text{ at rate }J(N, \phi)\\
\text{Maturation : } (N_{A},N_{B}) \rightarrow & (N_{A}-1,N_{B}+1) \text{ at rate }k_{m}(\phi)N_{A}  \\
\text{Budding : } (N_{A},N_{B}) \rightarrow & (N_{A},N_{B}-n_v)\text{ at rate }K(\phi)N_{B} \quad  {\rm if}\  N_B\geq n_v \\
\text{  } (N_{A},N_{B}) \rightarrow & (N_{A},0)\text{ at rate }K(\phi)N_{B} \quad  {\rm if}\  N_B<n_v \\
\end{split}
\label{dyneqs_stochastic}
\end{equation}
The compartment also contains a neutral species contributing to its total size $N$. Hence, for completeness, we need to explicitly define the dynamics of $N$:
\begin{equation}
\begin{split}
\text{Fusion : } N \rightarrow & N+N_v \text{ at rate }J(N, \phi)\\
\text{Budding : } N \rightarrow & N-N_v \text{ at rate }K(\phi)N_{B} \quad  {\rm if}\  N_B\geq n_v \\
\text{  } N \rightarrow & N-N_v \text{ at rate }K(\phi)N_{B} \quad  {\rm if}\  N_B<n_v \\
\end{split}\label{e2}
\end{equation}
From this set of stochastic equations, a first step towards constructing a mean-field approximation is to compute the time derivative of the average quantities $N$,$N_A$ and $N_B$:
\begin{equation}
\begin{split}
\langle \dot{N} \rangle = & N_v\langle J(N,\phi)\rangle-N_v\langle K(\phi)N_B\rangle \\
\langle \dot{N_A} \rangle = & n_v\langle J(N,\phi)\rangle -\langle k_m(\phi)N_A \rangle \\
\langle \dot{N_B} \rangle = & \langle k_m(\phi)N_A \rangle - n_vP_{N_B\geq n_v}\langle K(\phi)N_B\rangle_{N_B\geq n_v} -P_{N_B\geq n_v} \langle K(\phi) {N_B}^{2}\rangle_{N_B<n_v} \\
\end{split}\label{exact}
\end{equation}
Where, $\forall t \geq 0$, $\langle X \rangle_{Y}$ is the (ensemble) average value of $X(t)$ given constraint $Y$, and $P_Y$ is the probability that constraint $Y$ is verified at time $t$. It is important to note \eq{exact} is a direct consequence from \eq{dyneqs_stochastic} and \eq{e2} and is therefore exact. Now to obtain a mean-field approximation, one traditionally replaces every average of products or non-linear combinations of variables by products or combinations of the averages. Also fluctuations are neglected so $P_{N_B\geq n_v}$ is taken equal to $1$ when $\langle N_B \rangle \geq n_v$. This leads to the following complete mean-field equation:
\begin{equation}
\begin{split}
\dot{N}= & N_v(J(N,\phi)-K(\phi)N_B) \\
\dot{N_A}= & n_vJ(N,\phi)-k_m(\phi)N_A \\
\dot{N_B}= & k_m(\phi)N_A - n_vK(\phi)N_B\text{ if }N_B\geq n_v \\
\dot{N_B}= & k_m(\phi)N_A - K(\phi){N_B}^2 \text{ if }N_B< n_v \\
\end{split}
\end{equation}
Because of the change of behavior when $N_B<n_v$, the general solution of this system is complex. The mean-field equations are generally valid for large systems when fluctuations can be neglected. In such cases we can assume that $N_B\geq n_v$. In this case we can easily relate $\dot{N}$ and $\dot{N_A}+\dot{N_B}$:
\begin{equation}
\frac{\dot{N}}{N_v}=\frac{\dot{N_A}+\dot{N_B}}{n_v}
\end{equation}
From this we conclude that the relation $(N_A+N_B)/N=n_v/N_v$ is valid for all time, provided that it is verified at $t=0$. This allows to derive the self-consistent equations for the mean-field steady-state:
\begin{equation}
\frac{N}{N_v}=\frac{N_A+N_B}{n_v}=\frac{J(N,\phi)}{(1-\phi)k_m(\phi)}\qquad  \phi=\frac{k_m(\phi)}{k_m(\phi)+K(\phi)n_v}
\end{equation}
This corresponds to Eq.4 of the main text.

\section{First passage time for a deterministic maturation process}
In the main text, maturation of a single site is assumed to be a single step process. In reality the maturation mechanism might involve several steps. As a consequence the distribution of the maturation time of a given component would be different from a single exponential. The limit where maturation involves a large number of independent steps can be discussed relatively easily. Because of the central limit theorem, the distribution of the total maturation time of a single component is expected to be a Gaussian with a relatively small standard deviation. The extreme case is when the number of steps goes to infinity and the maturation is deterministic. We can compute the first passage time to full maturation of the compartment in this case. We choose the value $1/k_m$ for the deterministic maturation time of a single component in order to compare with the result of the main text. Here each site of type $A$ has a lifetime $1/k_m$, therefore full maturation occurs only if no new vesicle is injected during a time $1/k_m$ after the injection of the last vesicle. Vesicles are injected at a rate $J$ in the simplified model, and each injection completely resets the maturation state of the system. Since vesicle injections are independent, this leads to this expression for the average maturation time $\tau$:
\be
\tau=\frac{1}{k_m}+\langle n_{inj} \rangle \langle t_{inj} \rangle
\label{tau_def}
\ee
where $\langle n_{inj} \rangle$ is the average number of vesicle injections in addition to the first one and $\langle t_{inj} \rangle$ is the average waiting time between two injections. In order to compute these two quantities, we define the survival probability $p_S(t)$ which is the probability to reach the time $t$ without injecting a vesicle. The equation satisfied by $p_S(t)$, and its solution, are:
\begin{equation}
\frac{dp_S}{dt}=-Jp_S(t)\ ,\qquad p_S(t)=e^{-Jt}
\end{equation}
The probability of reaching $t=1/k_m$ without experiencing a vesicle injection (a reset) is $p_0\equiv p_S(1/k_m)$, hence the probability of experiencing exactly $n$ jumps before complete separation is $p_n=p_0(1-p_0)^n$. Consequently, the mean number of jumps is
\be
\langle n_{inj}\rangle=\sum_{n=1}^\infty n p_n=\frac{1-p_0}{p_0}
\label{avg_jump}
\ee
For one reset event, the probability density of the reset time is $-\frac{dp_S}{dt}$ so the average waiting time before a vesicle injection is given by:
\begin{equation}
\langle t_{inj}\rangle = \frac{\displaystyle \int_{0}^{1/k_m}t\frac{dp_S}{dt}dt}{\displaystyle \int_{0}^{1/k_m}\frac{dp_S}{dt}dt}=\frac{\displaystyle -p_0/k_m+\int_{0}^{1/k_m}p_S(t)dt}{1-p_0}
\label{avg_tjump}
\end{equation}
Introducing Eqs.(S\ref{avg_jump},S\ref{avg_tjump}) in \eq{tau_def}, we get the expression of the separation time:
\begin{equation}
\tau=\frac{1}{p_S(1/k_m)}\int_{0}^{1/k_m}p_S(t)dt=\frac{e^{\frac{J}{k_m}}-1}{J}
\end{equation}
This is exactly the same result as in the main text where maturation is made of a single step (Eq.(11)), which suggests that as long as the injection of new vesicle can be treated as a Poisson process, the first passage time to full maturation should not be affected much by the actual distribution of the maturation time of single components.

\section{Simulation scheme}
\subsection{General framework}
We make use of the Gillepie algorithm \cite{gillespie:1977} to perform exact stochastic simulations of our system. The Gillespie algorithm is a general method that allows to compute statistically correct trajectories for a stochastic system. The system must be described by a discrete set $S$ of states. Being in a given state $i\in S$ at time $t$, the system must have a probability $R_{i\rightarrow j}dt$ to jump from $i$ to another state $j$ between $t$ and $t+dt$. This definition is rather general, which allows the Gillespie algorithm to be applied in a variety of Physical contexts. Here the state space $S$ is composed of all the possible configurations of the compartment. A single state is  defined as a couple $(N_A,N_B)$ (or equivalently $(N,\phi)$ with $N=N_A+N_B$ and $\phi=N_B/N$) in which $N_{A}$ and $N_{B}$ are the numbers of A sites and B sites in the compartment. Below is the list of all the allowed transitions between states together with their transition rates:
\begin{equation}
\begin{split}
\text{Fusion : }(N_{A},N_{B}) \rightarrow & (N_{A}+n_v,N_{B})\text{ at rate }J(N,\phi) \\
\text{Maturation : } (N_{A},N_{B}) \rightarrow & (N_{A}-1,N_{B}+1)\text{ at rate }k_{m}(\phi)N_{A}  \\
\text{Budding : } (N_{A},N_{B}) \rightarrow & (N_{A},N_{B}-\text{min}(n_v,N_B))\text{ at rate }K(\phi)N_{B}  \\
\end{split}
\end{equation}
Where the functions $J(N,\phi)$, $k_m(\phi)$ and $K(\phi)$ can be arbitrary but must remain positive. After defining the system and its dynamics, the Gillespie algorithm  can be employed to generate statistically exact trajectories using the following scheme:
\begin{enumerate}
\item Initialize a random number generator.
\item Initialize the system at $t=0$ in a state $\Omega$. Here we choose the state $\Omega =(n_v,0)$, meaning that we start with a single vesicle of type $A$ as a compartment.
\item Compute the sum $\Sigma$ of all the transition rates out of the state $\Omega$, and use it to generate the waiting time $\Delta t$ before the next event using the distribution $\Sigma e^{-\Delta t \Sigma}$.
\item Choose randomly the state $\Omega'$ reached after $\Delta t$ from the list of possible states, knowing that the probability of having $\Omega \rightarrow \Omega'$ is given by $R_{\Omega\rightarrow \Omega'}/\Sigma$.
\item Update the time to $t+\Delta t$ and the state $\Omega$ accordingly with whatever event has been selected.
\item Loop back to step $3$.
\end{enumerate}

This scheme has been proved to generate trajectories that are exact in the sense that the probability for the program to generate a given trajectory is equal to the probability of observing this trajectory in the real system. In this paper we implemented this algorithm in the C language using the Mersenne-Twister pseudo random number generator.

\subsection{Construction of the phase diagram}
We aim at computing the value of the parameter $\eta$ defined in the main text. In order to compute this quantity we need to obtain average values for $N_{ves}$ the total number of emitted vesicles and $N_{mat}$ the final size of the compartment after complete maturation. We therefore performed a large number of simulations all identical apart from the initial state of the random number generator. We initialize the system at $t=0$ with $(N_{A},N_{B},N_{ves})=(n_v,0,0)$ and let the simulation run. After each time step we increase $N_{ves}$ if a vesicle budding has happened and check for complete maturation which is defined by $N_{A}=0$. Once complete maturation or a maximum number of simulation steps is reached the simulation is stopped. $N_{ves}$ is registered together with the final size $N_{mat}$ of the compartment. We perform 320 of these simulations (for each set of parameters) and we require that at least 90\% of them went to full maturation to compute satisfying estimates of the average values $\langle N_{ves}\rangle$ and $\langle N_{mat}\rangle$ (see error bars on Fig.3 - main text). This gives the value of $\eta$ for a specific value of $J$, $k_{m}$ and $K$, which corresponds to one point in the phase diagram of Fig.3 in the main text. The process is then repeated in an automated way to compute the value of $\eta$ for all the 20x20 points. For certain points (small values of $\phi_B$ and/or large values of $N$), less than 90\% of the simulations get to full maturation. We therefore cannot precisely compute $\eta$ but we verified that we always had $\eta>0.99$. The value of $\eta$ displayed on Fig.3 for these points is arbitrarily set to 1.

\section{Additional results}

\subsection{Phase diagram for constant rates}
\begin{figure}[htbp] 
   \centering
   \includegraphics[width=15cm]{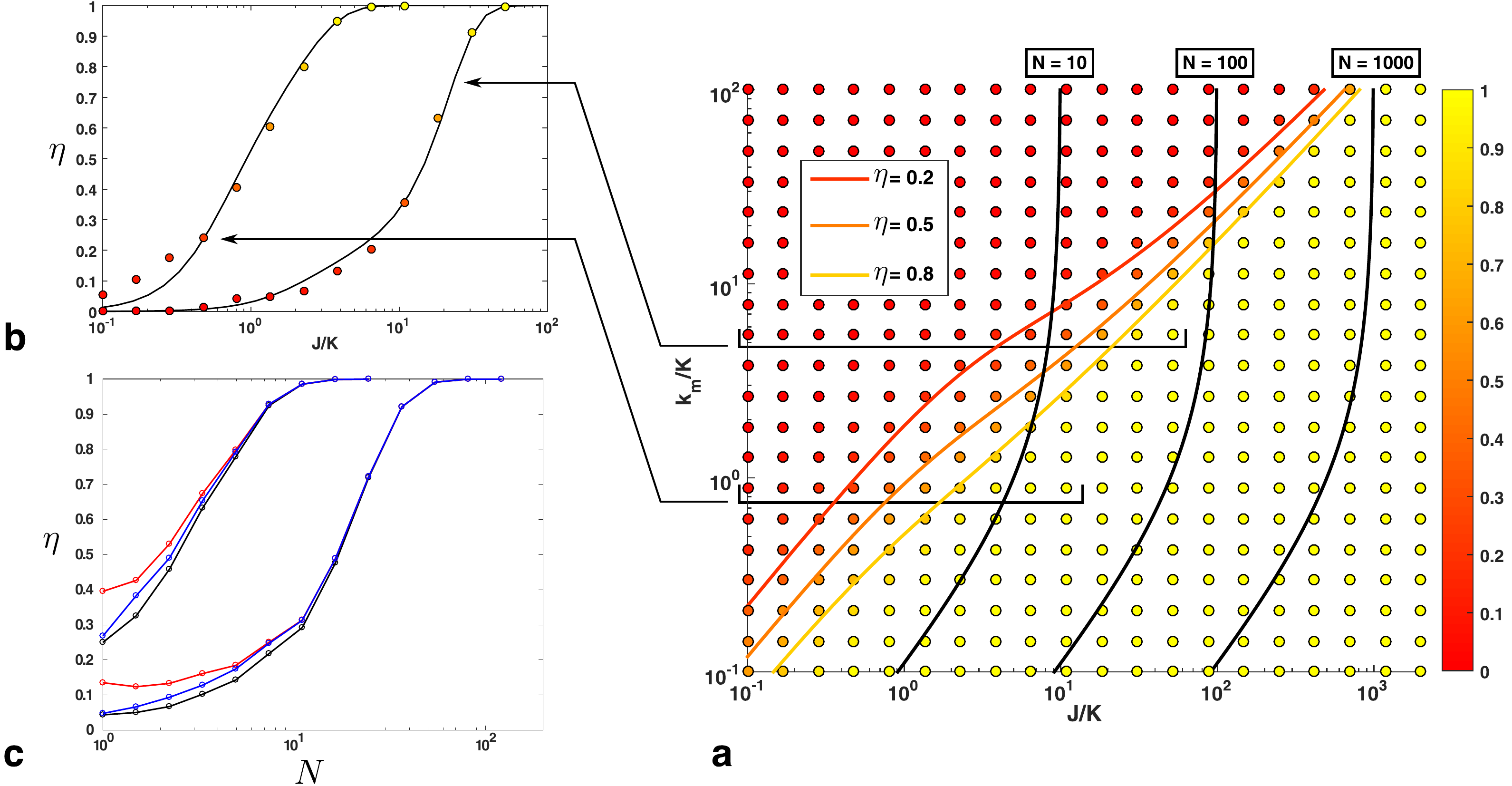}
   \caption{Simulation results. {\bf a}. Phase diagram: the value of the output parameter $\eta$ as a function of the in-flux $J$ and maturation rate $k_m$, normalized by the budding rate $K$, illustrates the transition between the {\em vesicular exchange} ($\eta\simeq1$) and {\em compartment maturation} ($\eta\simeq0$) regimes. The black lines show constant values of the compartment size in the pseudo steady-state (Eq.4 - main text). The three colored lines represent constant values of $\eta$ as given by the approximate analytical computation of Eq.12 - main text. {\bf b}. Cuts through the phase diagram varying the compartment size for two fixed pseudo steady-state compositions. {\bf c}. Output parameter as a function of the steady-state compartment size for two values of the steady-state composition $N_B/N_A$, as in Fig.2c of the main text. The black curves includes all possible events and is identical to the curves of Fig.2c. For the red curve, all direct maturation of a compartment, prior to any fusion event, have been removed. For the blue curve, all events where the fully matured compartment is of unit size, either because of direct maturation or after fusion and budding events, have been removed.}
   \label{FigSI_phasediag}
\end{figure}

In Fig.2 of the main text the phase diagram for the transition between the vesicular export dominated regime and the compartment maturation dominated regime is presented as a function of the quasi-steady-state values of the compartment size and composition. This is to ease the comparison between the different models, for which a given steady-state corresponds to different values of the kinetic rates. For completeness, we present in \fig{FigSI_phasediag} the same phase diagram as a function of the model parameters $J/K$ and $k_m/K$.\\
As discussed in the text, the size distribution of mature compartment presents a peak at small size in the basic model. This corresponds to the direct maturation of small compartments with a size comparable to that of a single vesicle. This contribution is due to the fact that our model does not include any specific process for the nucleation of a compartment and the initiation of maturation. To see the influence of direct maturation on the output parameter $\eta$, we present in \fig{FigSI_phasediag}c modified output parameters where direct maturation events have been removed, and where all events for which the fully matured compartment is of unit size (because of direct maturation or following several fusion and budding events), have been removed. It shows that these types of events do not qualitatively modify the picture, and make a quantitative difference only for very small steady-state compartment size.

\subsection{Scatter plots for the outflux}
\begin{figure*}
\centering
\includegraphics[width=\textwidth]{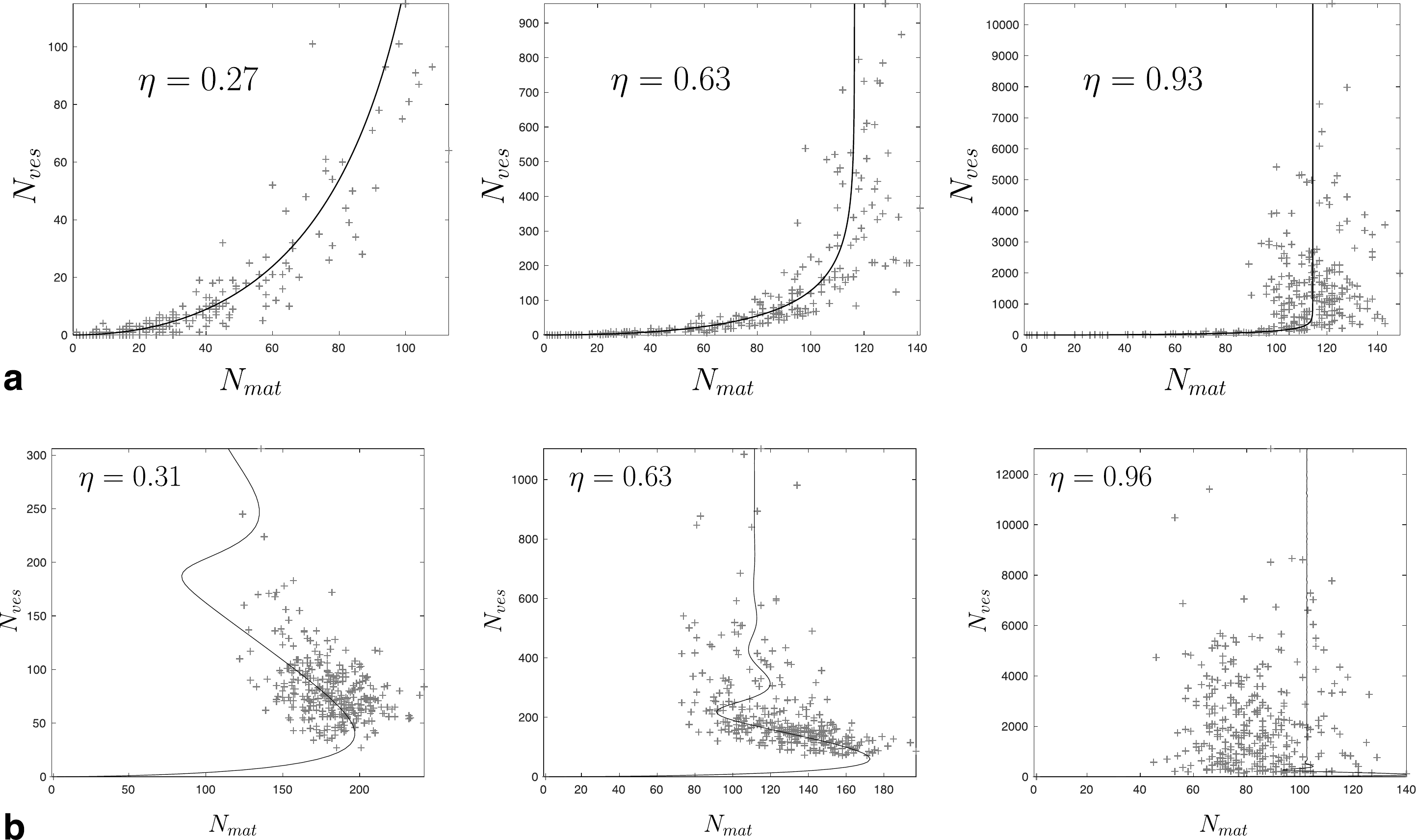}
\caption{\label{fig:scatter} Scatter plots for the distributions of $N_{ves}$ and $N_{mat}$ (crosses) obtained from 320 independent simulations (for three different values of $\eta$) for {\bf a.} a constant influx and {\bf b.} a composition-dependent influx $J\sim(1-\phi)$. The solid lines are analytical estimates obtained with \eq{eq:scatter}.}
\end{figure*}

The output parameter $\eta$ is computed using the average of $N_{ves}$ and $N_{mat}$ over many simulations, and therefore brings no information on the fluctuations of these quantities. These fluctuations are shown as scatter plots of $N_{ves}$ and $N_{mat}$ on \fig{fig:scatter}a for constant rates and  \fig{fig:scatter}b with homotypic fusion. The distribution of the values can be understood using mean-field arguments. Assuming that full maturation occurs after a time $t_{mat}$, One may defined $N_{ves}$ and $N_{mat}$ as:
\begin{equation}
N_{ves}=\int_{0}^{t_{mat}}KN_B(t)dt \qquad N_{mat}=N_{B}(t_{mat})\label{eq:scatter}
\end{equation}
where $N_B(t)$ is assumed to follow the mean-field evolution given by \eq{dyneqs}.
This estimate, shown as solid lines in \fig{fig:scatter}, accurately reproduces the observed distribution. This suggests that, for the parameters of \fig{fig:scatter}, the main source of fluctuation comes from the distribution of the isolation time $t_{mat}$.

\subsection{Role of homotypic fusion: Linear stability analysis of the steady-state}

The mean-field equations, in the presence of homotypic fusion and cooperative maturation, read:
\begin{eqnarray}
\dot A=n_v J(\phi)-J_m\quad {\rm and}\quad \dot B=J_m-n_vKB\cr
{\rm with} \quad J(\phi)=J_0(1-\phi)\quad {\rm and }\quad J_m=k_mA(1+\alpha\phi)
\label{dyneqs}
\end{eqnarray}
with $\phi=\frac{B}{A+B}$. 

In the following, we normalise rates by the budding rate: $j_0=J_0/K$ and $\bar k_m=k_m/(n_vK)$. One fixed point of these equations is $A=B=0$, the other satisfies the equations:
\be
A=\frac{B^2}{j_0-B}\qquad \bar k_m\alpha B^2-(1+\bar k_m(1+\alpha))j_0B+j_0^2=0
\ee
A non-zero fixed point always exists, given by:
\be
A=\frac{B^2}{j_0-B}\qquad B=\frac{j_0}{2\bar k_m\alpha}\left(1+\bar k_m(1+\alpha)-\sqrt{1+2\bar k_m(1-\alpha)+\bar k_m^2(1+\alpha)^2}\right)
\ee

This fixed point is always stable, but one can show that the kinetics toward the fixed point is overdamped only if
\begin{eqnarray}
-2 (\alpha +1)^3\bar k_m^3+\left(5 \alpha ^2+2 \alpha -2\right)\bar k_m^2-4 \alpha \bar k_m+1>\hspace{5cm}\cr
2\bar k_m \left((\alpha +1)^2\bar k_m-\alpha \right) \sqrt{\bar k_m \left(-2 \alpha +(\alpha +1)^2
  \bar k_m+2\right)+1}
   \end{eqnarray}

\begin{figure}[htbp] 
   \centering
   \includegraphics[width=15cm]{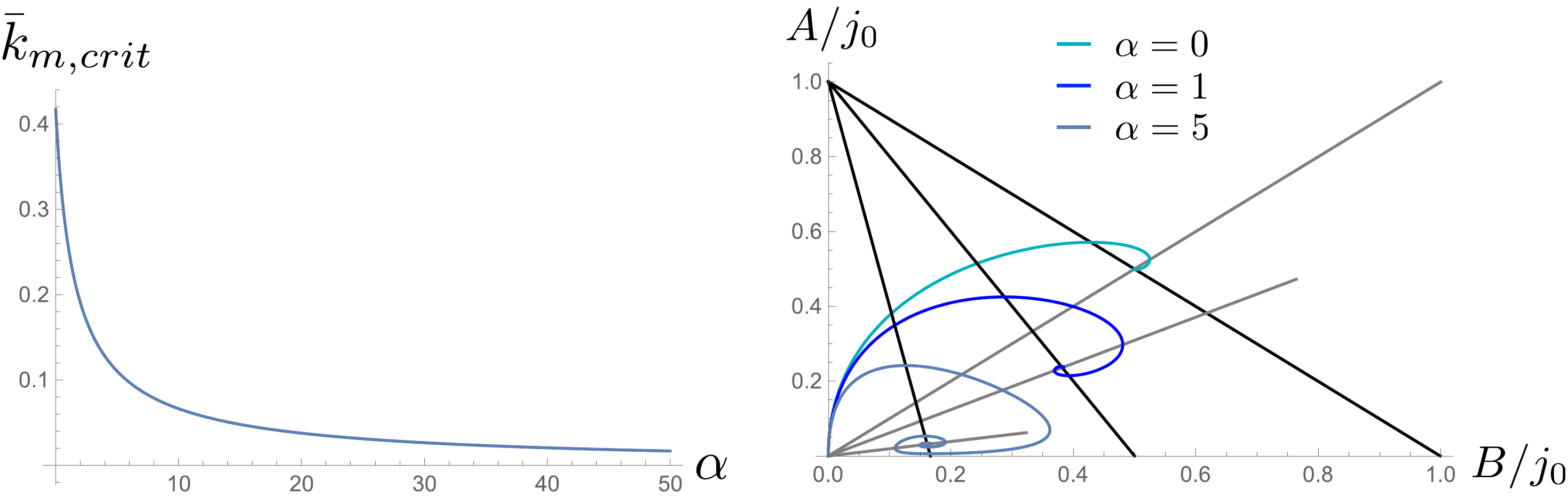}
   \caption{Left, Value of the critical maturation rate $\bar k_{m,{\rm crit}}$ beyond which the system shows damped oscillations around the fixed point. Right. Trajectories in the phase space $\{A(t),B(t)\}$ for $\bar k_m=1$ and different values of $\alpha$ ($=0,\ =1,\ =5$). The black (gray) lines are the corresponding null clines $\dot A=0$ ($\dot B=0$).}
   \label{FigSI_linear_stability}
\end{figure}

This means that there is a decreasing function $\bar k_{m,\rm{crit}}(\alpha)$ (shown in \fig{FigSI_linear_stability}, and whose maximum is $\bar k_{m,\rm{crit}}(0)\simeq 0.42$) such that when $\bar k_m>\bar k_{m,\rm{crit}}(\alpha)$ the system reaches the steady-state undergoing oscillations. The larger $\bar k_m$, the closer the oscillations bring the system to the full maturation adsorbing boundary $A=0$. The full maturation probability thus becomes relatively independent on the system's size in that case. Note that this behaviour rely mostly on the existence of homotypic fusion, and can in principle be observed in the absence of cooperativity $\alpha=0$, although cooperativity does decreases the threshold $\bar k_{m,crit}$ beyond which relaxation is underdamped.

\subsection{Phase diagrams for the full model.}
\begin{figure}[htbp] 
   \centering
   \includegraphics[width=15cm]{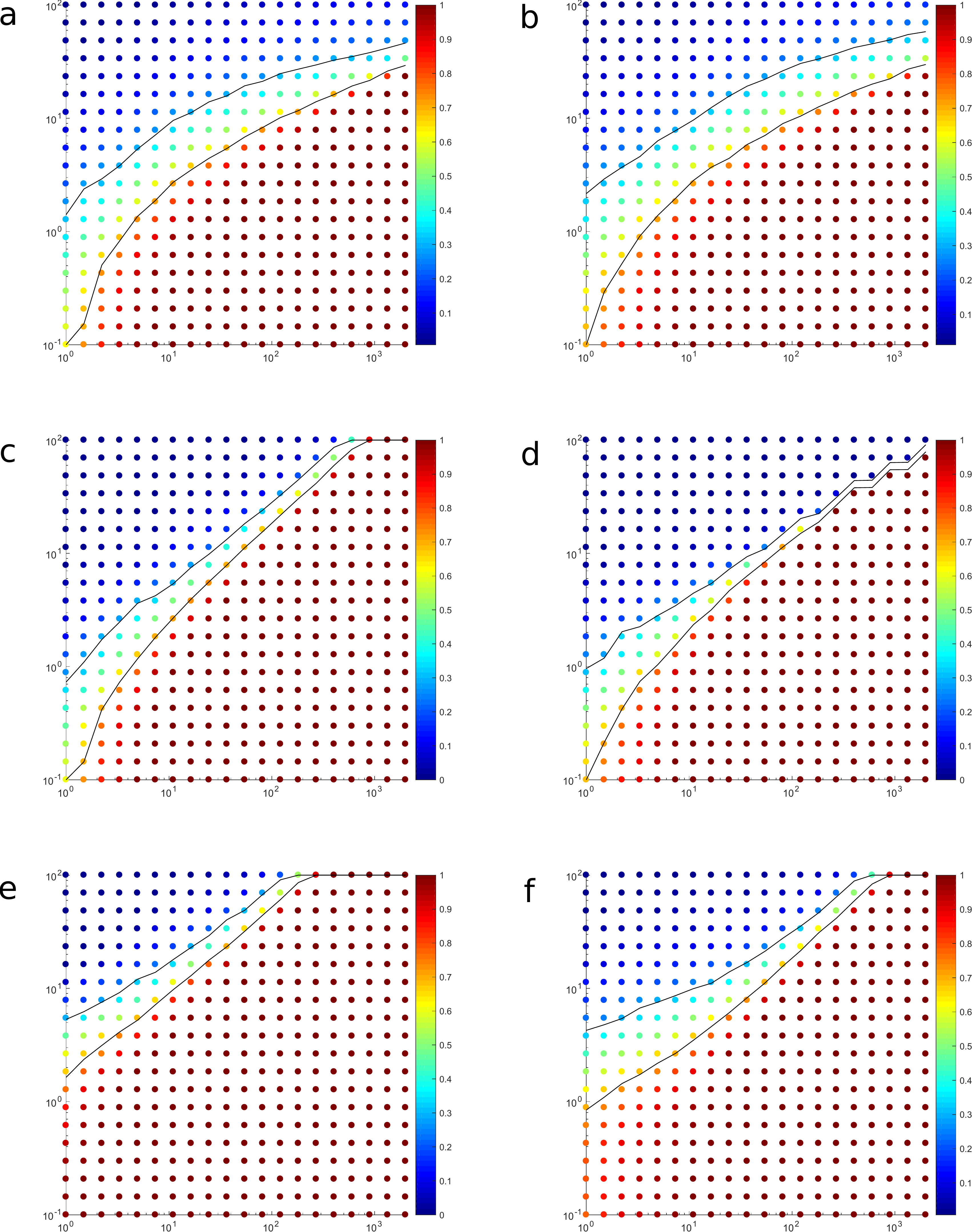}
    \caption{Phase diagram for the output parameter $\eta$ as in Figs.2-3 of the main text, for different values of the parameters of the full model: {\bf a}. $J=J_0(1-\phi)$, $n_v=1$ {\bf b}. $J=\sqrt{N}J_0(1-\phi)$, $n_v=1$ {\bf c}. basic model, {\bf d}. $J=J_0\sqrt{N}$, $n_v=1$ {\bf e}. $n_v=10$ and {\bf f}. $k_m(\phi)=k_m(1+\alpha\phi)$ with $\alpha=50$, $n_v=1$.}
   \label{FigSI_phase_diag}
\end{figure}
In the full model, the fluxes characterising the influx of immature ($A$) components, their maturation into mature ($B$) components, and the exit of $B$ components, are written, respectively:
\begin{eqnarray}
J_{A,in}=J(N, \phi)n_v\quad,\quad J_{A\rightarrow B}=k_m(\phi)N_A\quad,\quad J_{B,out}=K(\phi) n_v N_B\cr
J(N,\phi)=N^\beta(1-\phi)\quad,\quad k_m(\phi)=k_m(1+\alpha\phi)\quad,\quad K(\phi)=K\phi
\label{fluxes}
\end{eqnarray}
where the parameter $\beta$ quantifies the size dependence of the influx and the parameter $\alpha$ quantifies the cooperativity of the maturation process. \fig{FigSI_phase_diag} shows different phase diagrams varying these parameters. It complements the Fig4 of the main text, which only shows transition boundaries for  $\eta-0.5$. In addition, it explores the case where the influx depends on the size of the compartment: $\beta=1/2$, which corresponds to a diffusion-limited fusion of the vesicular influx. The main result is that the width of the transition is mostly influenced by the mechanism of homotypic fusion. The transition is  smoother when homotypic fusion is treated in a continuous manner ($J\sim(1-\phi)$ - \fig{FigSI_phase_diag}a,b). This reflects the fact that in the former case, full compartment maturation is an almost deterministic process that occurs during the first oscillation, hence lacks the exponential nature related to stochasticity. One the other hand, a size dependent influx makes little difference to the phase diagram (compare  \fig{FigSI_phase_diag} a and b, or  \fig{FigSI_phase_diag}  c and d).

\addcontentsline{toc}{chapter}{Supporting references}